%% file: paper.tex
\documentclass[twocolumn,showpacs,aps,prl,superscriptaddress,nofootinbib]{revtex4}

\usepackage{graphicx}

\usepackage{dcolumn}

\usepackage{epsfig}

\input babarsym

\def\bldmth{}
\def\beq{\begin{equation}}
\def\eeq{\end{equation}}

\def\beqa{\begin{eqnarray}}
\def\eeqa{\end{eqnarray}}

\def\RE{ROE}
\def\asymvalue{0.02 \pm 0.16}
\def\asymvaluesys{0.02 \pm 0.16~{\rm(stat.)} \pm 0.03~{\rm(syst.)}}
\def\brvalue{(5.5 \pm 1.0)\times 10^{-6}}
\def\brvaluesys{(5.5 \pm 1.0~{\rm(stat.)} \pm 0.7~{\rm(syst.)})\times 10^{-6}}

\def\DKgood{{\ensuremath{DK_D}}}
\def\DKbad {{\ensuremath{DK_{\not D}}}}

\def\Dpigood{{\ensuremath{D\pi_D}}}
\def\Dpibad {{\ensuremath{D\pi_{\not D}}}}
\def\Dpix   {{\ensuremath{D\pi X}}}
\def\DpiX   {\Dpix}
\def\DKX    {{\ensuremath{DKX}}}
\def\BBgood {{\ensuremath{BBC_D}}}
\def\BBbad {{\ensuremath{BBC_{\not D}}}}
\def\qqgood {{\ensuremath{qq_D}}}
\def\qqbad {{\ensuremath{qq_{\not D}}}}
\def\NBB {{\ensuremath{N_{{BB}_{\not D}}}}}

\def\DE{\ensuremath{\Delta E}}
\def\mes{\ensuremath{m_{ES}}}
\def\mD{\ensuremath{m_D}}
\def\NNqq{\ensuremath{q}}
\def\NNcomb{\ensuremath{d}}

\def\pB{\ensuremath{\mathbf{p_B}}}

\def\P{\ensuremath{{\cal P}}}  
\def\E{\ensuremath{{\cal E}}}  
\def\Q{\ensuremath{{\cal Q}}}  
\def\C{\ensuremath{{\cal D}}}  
\def\B{{\cal B}}   

\def\A{{{ A}}}  

\def\btou{\ensuremath{b\to u\cbar s}}
\def\btoc{\ensuremath{b\to c\ubar s}}
\def\ppp{\ensuremath{\pi^+\pi^-\pi^0}}

\def\btodzerok{\ensuremath{B^- \to \Dz K^-}} 
\def\btodkgen{B\to D^{(*)0}K^{(*)}}
 
\def\dtoppp{\ensuremath{D\to \ppp}}
\def\dztoppp{\ensuremath{\Dz\to \ppp}}
\def\dztokpp{\ensuremath{\Dz\to K^-\pi^+\piz}}
\def\Dppp{\ensuremath{D_{\pi^+\pi^-\piz}}}
\def\decaychain{\ensuremath{B^- \to \Dppp K^-}}

\def\bMtodpi{\ensuremath{B^- \to D^0 \pi^-}} 
\def\bMtodpiKpp{\ensuremath{\bMtodpi, \ \dztokpp}}
\def\bMtodpippp{\ensuremath{\bMtodpi, \ \dztoppp}}

\def\BR#1{\ensuremath{{\cal B}(#1)}}

\newcommand{\BABARPubNumber}  {05/016}

\newcommand{\SLACPubNumber} {11245}

\def\figurebox#1#2#3{%
    \def\arg{#3}%
    \ifx\arg\empty
    {\hfill\vbox{\hsize#2\hrule\hbox to #2{\vrule\hfill\vbox to #1{\hsize#2\vfill}\vrule}\hrule}\hfill}%
    \else
    {\hfill\epsfbox{#3}\hfill}%
    \fi}

\begin{document}

\preprint{\BABARPubNumber} 
\preprint{SLAC-PUB-\SLACPubNumber} 

\begin{flushleft}
\babar-PUB-\BABARPubNumber\\
SLAC-PUB-\SLACPubNumber\\
\end{flushleft}

\title{\large \bf 
Measurement of the branching fraction and decay rate asymmetry of
\boldmath $\decaychain$
}

\input authors_apr2005.tex

\date{\today}

\begin{abstract}
We report the observation of the decay $\decaychain$,
where $\Dppp$ indicates a neutral $D$ meson detected in the final
state \ppp, excluding $\KS\piz$.
This doubly Cabibbo-suppressed decay chain can be used to measure the
CKM phase $\gamma$.
Using about $229$ million $\epem\to\BB$ events recorded by the
\babar\ experiment at the \pep2\ $\epem$ storage ring, we measure the
branching fraction $\B(\decaychain) = \brvaluesys$ and the decay rate
asymmetry  $\A = \asymvaluesys$
for the full decay chain.

\end{abstract}
\pacs{13.25.Hw, 12.15.Hh, 11.30.Er}

\maketitle
\vskip .3 cm



The Cabibbo-Kobayashi-Maskawa (CKM) matrix, whose element 
$V_{ij}$~\cite{ref:km} describes the weak charged-current 
coupling between
quark flavors $i$ and $j$, provides an explanation for CP~violation
in the Standard Model.
A crucial part of the program to study CP~violation
is the measurement of the angle $\gamma = \arg{\left(- V^{}_{ud}
V_{ub}^\ast/ V^{}_{cd} V_{cb}^\ast\right)}$ of the unitarity triangle
related to the CKM matrix.
The decays $\btodkgen$ can be used to measure $\gamma$ with
essentially no hadronic uncertainties, making use of interference
between \btou\ and \btoc\ decay amplitudes. A
number of variations on the original method~\cite{Gronau:1991dp} 
have been developed, and
some have been explored experimentally. Employing multiple methods
helps to resolve discrete ambiguities and decrease the experimental
error.

An important class of $\gamma$ measurement methods involves $\btodkgen$
with multi-body $D$ decays~\footnote{We use the symbol $D$ to indicate 
	any linear combination of a $\Dz$ and a $\Dzb$ meson state.}.
In this technique, $\gamma$ is extracted from an analysis of the $D$-decay
Dalitz plot, and ambiguities are resolved through interference
between several $D$ decay amplitudes~\cite{Giri:2003ty}.
Both Belle~\cite{Abe:2003cn} and \babar~\cite{Aubert:2004kv} have used
this method to obtain limits on $\gamma$ with the Cabibbo-favored
decay $D\to \KS \pi^+\pi^-$. 
The same approach can be carried out with multi-body final
states that are produced by singly Cabibbo-suppressed decay of both
$\Dz$ and $\Dzb$~\cite{Grossman:2002aq}.
While these modes yield much smaller event samples, their interfering
$\Dz$ and $\Dzb$ decay amplitudes have similar magnitudes.
Therefore, their overall sensitivity to $\gamma$ is a~priori expected 
to be similar to that of Cabibbo-favored $D$ decays, where the interfering
amplitudes typically have very dissimilar magnitudes.

Among the singly Cabibbo-suppressed modes, the decay $\dtoppp$ has a
relatively large branching fraction~\cite{ref:pdg} and a simple Dalitz
plot dominated by broad $\rho$ resonances~\cite{Frolov:2003jf}, making
it attractive for the measurement of $\gamma$ with this technique.  
Its major difficulty is the relatively small
signal-to-background ratio, which results mainly from the high
combinatorial background associated with $\piz$ reconstruction.
In this article, we describe an analysis procedure with which to extract
the $\decaychain$ signal for later use in a Dalitz plot analysis 
measurement of $\gamma$,
and report the measured branching fraction and decay rate asymmetry of
this decay chain. 
Our result excludes the decay mode $D\to\KS\piz$,
which is a previously-studied CP-eigenstate 
not related to the method of Ref.~\cite{Giri:2003ty}.

The decay rate asymmetry $A = (N^+ - N^-)/(N^+ + N^-)$, where $N^+$
($N^-$) is the number of signal $B^+$ ($B^-$) decays, depends on the
weak and strong phases of the $B$ decay, as well as the $\Dz$ and
$\Dzb$ decay rate and phase variation over the Dalitz plot. Its
magnitude is at most of order $2r_B$, where $r_B$, estimated to be
about 0.1~\cite{Giri:2003ty}, is the ratio between the magnitudes of
the interfering \btou\ and \btoc\ amplitudes.
Due to interference, the branching fraction $\BR\decaychain$
may differ from the product
$\B_{\rm prod}\equiv \B(\btodzerok)\times\B(\dztoppp) = 
	(4.1 \pm 1.6)\times 10^{-6}$~\cite{ref:pdg}
by up to about $2 r_B \B_{\rm prod}$.



The data used in this analysis were collected with the \babar\
detector at the \pep2\ energy-asymmetric \epem\ storage ring.  The data
consist of 207~fb$^{-1}$ collected on the $\Upsilon(4{\rm S})$
resonance (on-resonance sample), and 21~fb$^{-1}$ collected at an
$\epem$ center-of-mass (CM) energy approximately 40~\mev below the
resonance peak (off-resonance sample).
Samples of simulated events were analyzed with the same
reconstruction and analysis procedure. These include an $\epem\to\BB$ sample
about three times larger than the data; a continuum $\epem\to\qqbar$ sample,
where $q$ represents a $u$, $d$, $s$, or $c$ quark, with equivalent
luminosity similar to that of the data; and a signal sample about 200
times larger than what is expected in the data.
The \babar\ detector, as well as the methods used for charged
and neutral particle reconstruction and identification are described in detail in Ref.~\cite{ref:babar}.




We select events using criteria designed to maximize the signal
branching fraction sensitivity and the reliability of the maximum
likelihood fit procedure described below. 
To suppress the continuum background, we require the ratio $H_2/H_0$
of the 2nd to the 0th Fox-Wolfram moments~\cite{ref:R2}, computed
from the momenta of all charged particles and photon candidates not
matched to tracks, to be less than 0.50.
Charged kaon candidates are required to have a high quality particle
identification measurement and be identified using kaon
selection criteria that reduce the pion background to less than 3\%.
The measured energy of photon candidates is required to be at least 30~\mev.
Photon candidate pairs whose invariant mass is within 25~\mevcc\ of
the nominal $\piz$ mass~\cite{ref:pdg} are combined to make $\piz$
candidates, to which we perform a constrained-mass fit in order to
improve the $\piz$ energy and momentum resolutions.  Throughout 
this article, we use the symbol $\gamma_h$ to refer to the harder
(higher-energy) of the two photons constituting a $\piz$
candidate, and $\gamma_s$ to denote the softer (lower-energy) photon.

We select $\dtoppp$ candidate decays by requiring the $\ppp$ invariant
mass \mD\ to be between 1.830~\gevcc and 1.895~\gevcc. The \mD\ resolution
is about 14~\mevcc. The $D$ candidate energy and momentum resolutions
are then improved by performing a constrained mass fit.
The charged pion candidates are required to fail kaon selection
criteria.
The decay $D\to\KS\piz$ is rejected by excluding $\pi^+\pi^-$
candidate pairs whose invariant mass is between 0.489~\gevcc and 0.508~\gevcc.
We note that this last requirement will not be needed
when measuring $\gamma$ with an analysis of the
$\ppp$ Dalitz plot, where the $\KS\piz$ final state 
can be included as an incoherent term, as done in Ref.~\cite{Frolov:2003jf}.

Candidate \decaychain\ decays are constructed by combining a $\dtoppp$
candidate with a charged kaon candidate.
Additional continuum suppression is obtained by requiring
$|\cos\theta_T|<0.8$, where $\theta_T$ is the angle between the thrust
axis calculated in the CM frame with the daughters of the $B$
candidate and the thrust axis of the rest of the event (\RE).
For each $B$ candidate we calculate the beam-energy substituted mass
$\mes \equiv \sqrt{E_{\rm CM}^2/4 - |\pB|^2}$, where the total CM
energy $E_{\rm CM}$ is continuously determined from the measured \pep2\ beam
energies, and \pB\ is the momentum of the $B$ candidate in the CM
frame. Signal events have a Gaussian \mes\ distribution that peaks at the
nominal $B^-$ mass with a width of about 2.7~\mevcc, while 
background is distributed more broadly than signal. We require $5.272 <
\mes < 5.300$~\gevcc.
The energy difference $\DE = E_B - E_{\rm CM}/2$, where $E_B$
is the CM energy of the $B$ candidate, is required to be between 
$-70$~\mev\ and $60$~\mev. The \DE\ distribution of signal events peaks
around 0~\mev\ with a width of 21~\mev.

About 25\% of the events selected have more than one $B$ 
candidate. In these events, we select one $B$ candidate at random.
Random selection allows consistent studies of background suppression
variables, and degrades the signal sensitivity by only a few percent
relative to the best possible selection method.
%

Studying the simulated event sample selected by the above criteria, we
identify ten event types, one signal and nine background. We list
these types with the labels used to refer to them throughout the
article:
\begin{itemize}
\item {\bldmth \DKgood\ :} \decaychain\ events that were correctly 
reconstructed. These are the only events considered to be signal.

\item {\bldmth \DKbad\ :} \decaychain\ events in which the $D$ candidate is
misreconstructed, namely, some of the particles used to form the final
state $\ppp$ do not originate in the decay of the $D$ meson.

\item {\bldmth \Dpigood\ :} $\bMtodpi$, $D^0\to\ppp$ decays, 
where the decay $D^0\to \ppp$ is correctly reconstructed and the
remaining $\pi^-$ is mistaken to be the kaon. 

\item {\bldmth \Dpibad\ :} $\bMtodpi$, $D^0\to\ppp$ decays, where
the  $D$ candidate is misreconstructed. 
The kaon candidate may be either the remaining $\pi^-$ or a particle from
the other $B$ meson in the event.

\item {\bldmth \DKX\ :} $B\to D^{(*)} K^{(*)-}$, excluding
$\dtoppp$ decays, with a misreconstructed $D$ candidate.

\item {\bldmth \DpiX\ :} $B\to D^{(*)} \pi^-$ and $B\to D^{(*)}
\rho^-$, excluding $\dztoppp$ decays, with a misreconstructed $D$
candidate.

\item {\bldmth \BBbad\ :} All other $\BB$ events with a
misreconstructed $D$ candidate.  

\item {\bldmth \BBgood\ :} Other $\BB$ events with a correctly
reconstructed $\dtoppp$ decay.  

\item {\bldmth \qqbad\ :} Continuum $\epem\to\qqbar$ events with a
misreconstructed $D$ candidate.  

\item {\bldmth \qqgood\ :} Continuum $\epem\to\qqbar$ events with a correctly
reconstructed $\dtoppp$ decay.  

\end{itemize}
The Cabibbo-favored decay chain $B^-\to \Dz \pi^-$, $\Dz\to
K^-\pi^+\pi^0$, which has the same final state particles as our
signal decay, does not contribute significantly to the background,
since it is suppressed by the particle identification and \mD\ cuts


The majority of background events are of the \qqbad\ type.
The combination of \DpiX, \DKX, and \BBbad\ events constitutes
the second largest background.
In order to suppress these backgrounds, we have developed two neural
networks, each of which combines several input variables that provide 
separation between signal and background.
The first neural network variable $\NNqq$ is computed from input 
variables that provide separation between continuum and \BB\ events.
The second variable $\NNcomb$ combines input variables 
that separate correctly reconstructed $\piz$ and $\Dz$ candidates from
misreconstructed ones. It provides separation between signal and
all misreconstructed-$D$ background.

The input variables for \NNqq\ are 
(1) the cosine of the CM angle between \pB\ and the beams;
(2) $|\cos\theta_{T}|$;
(3-4) the zeroth and second Legendre moments of the momentum flow of the \RE\
	about the CM thrust axis of the $B$ candidate daughters;
(5) log of the distance along the beam direction
	between the reconstructed $B$ vertex and the  vertex of the \RE,
	computed as in Ref.~\cite{Aubert:2002ic};
(6) log of the distance of closest
	approach between the kaon track and the $D$ decay vertex, which is
	calculated from the $\pi^+$ and $\pi^-$ track parameters;
(7) an integer variable calculated from the probability that the \RE\
	contains a $\Bz$, determined using the lepton flavor tagging algorithm
	of Ref.~\cite{Aubert:2002ic}.

The input variables for \NNcomb\ are
(1) the invariant mass of the $\pi^0$ candidate;
(2) the $\pi^0$ momentum in the lab frame;
(3) cosine of the $\pi^0$ decay angle $\theta_{\piz}$, 
	defined as the angle between the $\gamma_h$ momentum and the 
	momentum of the CM frame, calculated in the $\piz$ rest frame; 
(4) the invariant mass $m_h$ of $\piz_h$,  where $\piz_h$ is the 
	$\piz$ candidate reconstructed from the $\gamma_h$ and any
	additional photon in the event except $\gamma_s$, 
	chosen such that $m_h$ is closest to the nominal $\piz$ mass;
(5) $m_s$, calculated analogously to $m_h$, but with $\gamma_s$
	instead of $\gamma_h$;
(6-7) cosines of the decay angles of the $\piz_h$ and the $\piz_s$, 
	calculated analogously to $\theta_{\piz}$;
(8) cosine of the 
	angle between $\pB$ and the thrust axis of the
	\ppp\ final state, calculated in the \ppp\ rest frame;
(9) cosine of
	the angle between the $D$ candidate momentum and the line 
	connecting the $B$ and $D$ decay vertices.

\begin{figure}[!htb]
\begin{center}
\begin{tabular}{cc} 
\includegraphics[height=4.1cm]{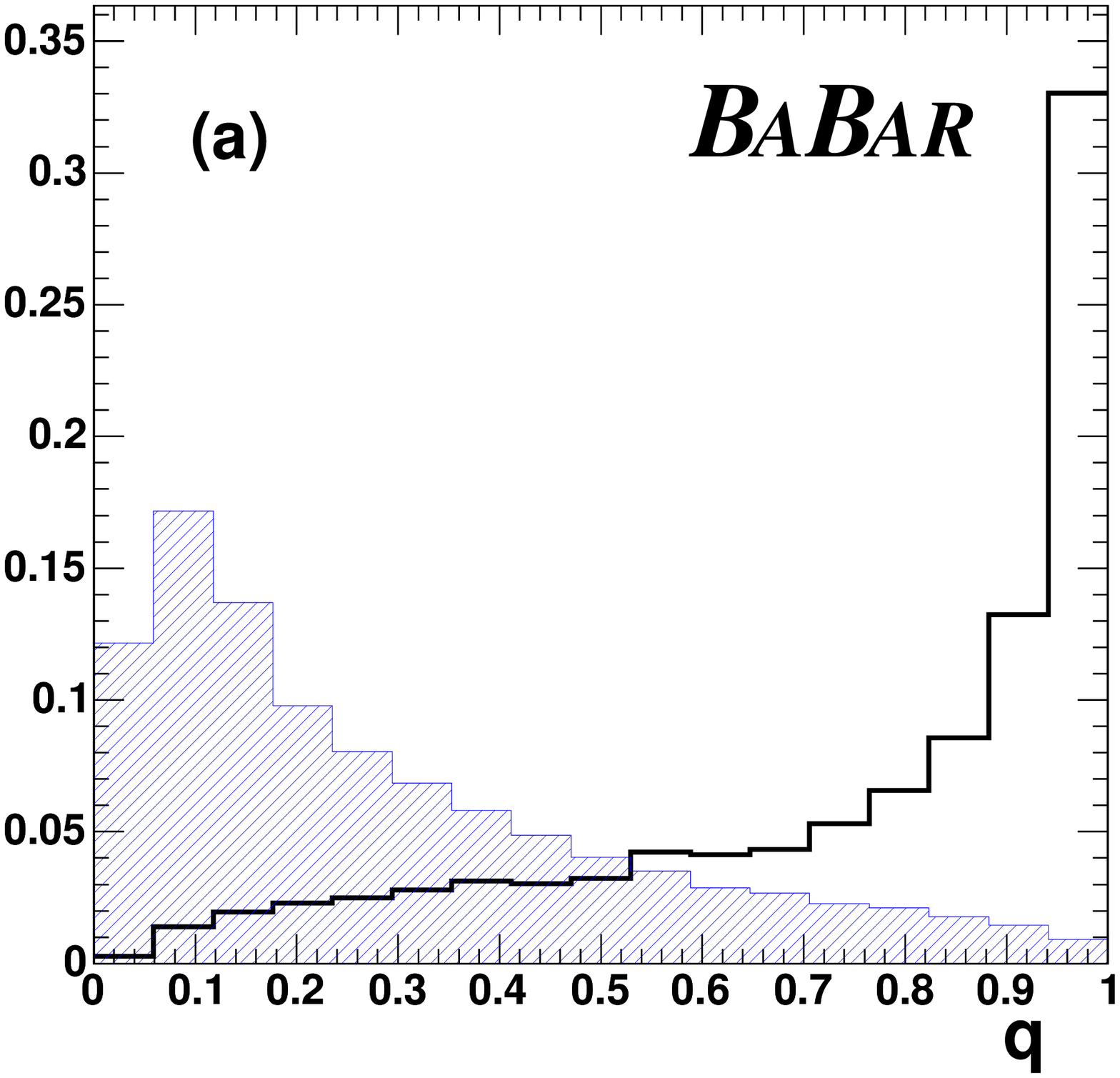}
&
\includegraphics[height=4.1cm]{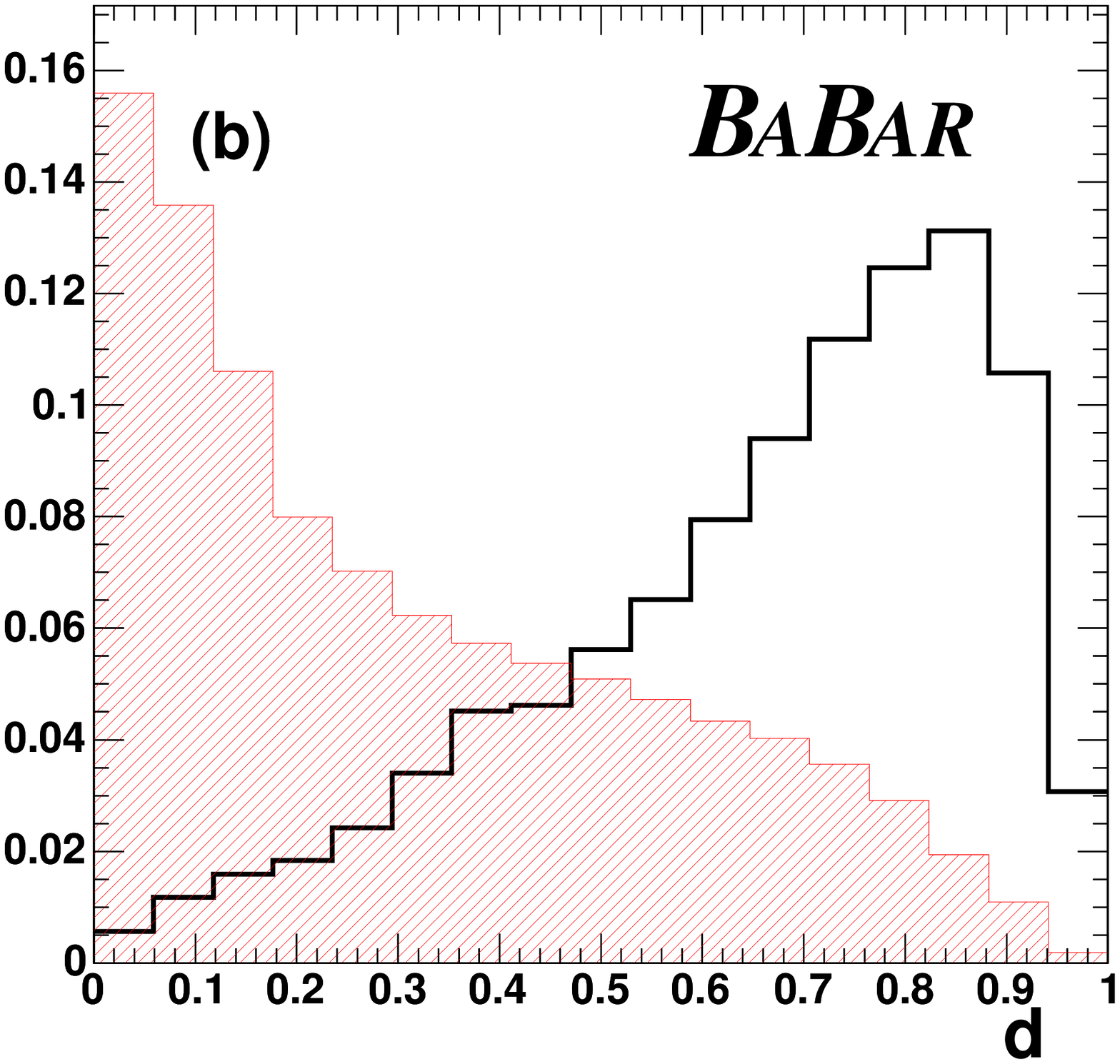} \\
\end{tabular}
\caption{(a) Distribution of the neural network variable \NNqq\ 
	for continuum (hatched) and signal simulated events. The 
	\BB\ background distribution is similar to that of signal.
	(b) Distribution of the neural network variable
	\NNcomb\ for \BB\ background
	(hatched) and signal simulated events. The continuum background 
	distribution is similar to that of the \BB\ background. 
	All histograms are normalized to unit area.}
\label{fig:NN}
\end{center}
\end{figure}

The \NNqq\ and \NNcomb\ distributions of simulated signal and
background events are shown in Fig.~\ref{fig:NN}. All events are
required to satisfy the conditions $\NNqq > 0.1$, $\NNcomb > 0.1$, in
order to reduce the background and suppress correlations
between the variables used in the fit described below.
The final signal reconstruction efficiency is 10.5\%.

We perform a maximum likelihood fit to measure the number 
and the decay rate asymmetry of signal
events in the on-resonance data sample, using the variables \DE,
\NNqq, and \NNcomb.
The variable \mes, which is commonly used as a fit variable in $B$
decay analyses, is not included in the fit. Studies with simulated
events indicate that correlations of \mes\ with other fit variables in
the distributions of \DpiX, \DKX, and \BBbad\ background events lead
to a bias in the measured signal yield, unless the correlations are
modeled correctly. Such modeling complicates the analysis procedure,
increases the dependence on the simulation, and incurs additional
systematic errors. By excluding
\mes\ from the fit, 
We give up some statistical precision in order to 
make the analysis more robust. Correlations between the \DE, \NNqq, and \NNcomb\
distributions for the different event types are at the few percent
level in the worst cases, and ignoring them in fits
to simulated events does not result in significant biases.

The probability density function (PDF) for the fit is
\beq
\P  = {1 \over \eta} \sum_{t} N_t \P_t(\DE, \NNqq, \NNcomb),
\eeq
where the subscript $t$ corresponds to one of the ten event types
listed above, $N_{t}$ is the number of events of type $t$, and
$\eta \equiv \sum_t N_t$.
The PDF $\P_{t}$ for events of type $t$ is a product of the form
\begin{equation} 
\P_{t}(\DE, \NNqq, \NNcomb) =
	  \E_t(\DE)\, \Q_t(\NNqq)\, \C_t(\NNcomb).
\label{eq:prodPDF}
\end{equation}
The functions $\E_\BBbad(\DE)$, $\E_\qqgood(\DE)$, and $\E_\qqbad(\DE)$
are parameterized as second order polynomials, and all other
$\E_t(\DE)$ functions are the sum of a Gaussian and a second order
polynomial.  The parameters of these functions are obtained from fits
to simulated events.
The PDFs $\Q_t(\NNqq)$ and $\C_t(\NNcomb)$ are 
15-bin histograms obtained from simulated events. 

To extract the signal yield and asymmetry, we minimize the log 
of the extended likelihood
\begin{equation}
 \calL = \frac{\eta^N e^{-\eta}}{N!}\prod^{N}_{i=1}\P(i).
 \end{equation}
Six parameters are floating in the fit. These are
the event yields
$N_\DKgood$,
$N_\Dpigood$, 
$N_\qqbad$, 
and $\NBB \equiv  N_\DKX + N_\Dpix + N_\BBbad$, 
the ratio $R_\Dpix \equiv N_\Dpix / \NBB$, 
and the decay rate asymmetry 
	$\A \equiv (N_\DKgood^+ - N_\DKgood^-) /
			  (N_\DKgood^+ + N_\DKgood^-)$,
where the superscript indicates the charge of the kaon. 
Five ratios of event yields are obtained from the simulation and 
are not varied in the fit. From these ratios we obtain the five parameters
	$N_\DKX    =  0.21   \, N_\DpiX       $, 
	$N_\Dpibad =  0.171  \, N_\Dpigood   $, 
	$N_\BBgood =  0.0089 \, \NBB        $, 
	$N_\qqgood =  0.0136 \, N_\qqbad    $, 
    and $N_\DKbad  =  0.1614 \, N_\DKgood   $. 
All fixed parameters are later varied to evaluate systematic errors,
as described below.

The results of the fit are summarized in Table~\ref{tab:results}.
We observe $N_\DKgood = 133 \pm 23$ signal events and 
the decay rate asymmetry $\A = \asymvalue$, 
where the errors are statistical only.
The corresponding branching fraction is 
$\B(\decaychain) = \brvalue$.
The statistical significance of the signal observation, obtained 
from a scan of the likelihood as a function of the signal yield, 
is 5.7 standard deviations.

\begin{figure}[!htpb]
\begin{center}
\begin{tabular}{cc}
\includegraphics[height=3.5cm]{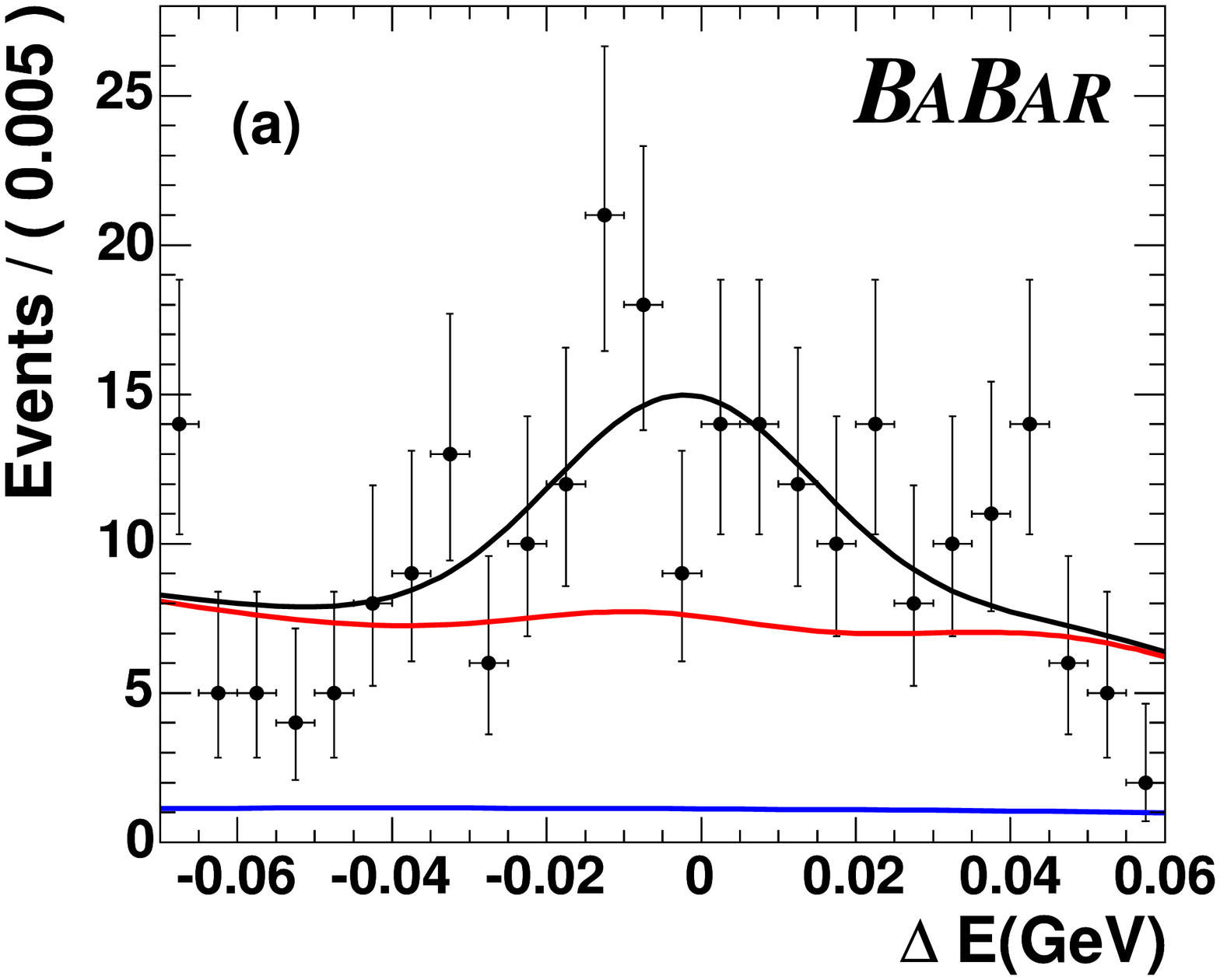} & 
\includegraphics[height=3.5cm]{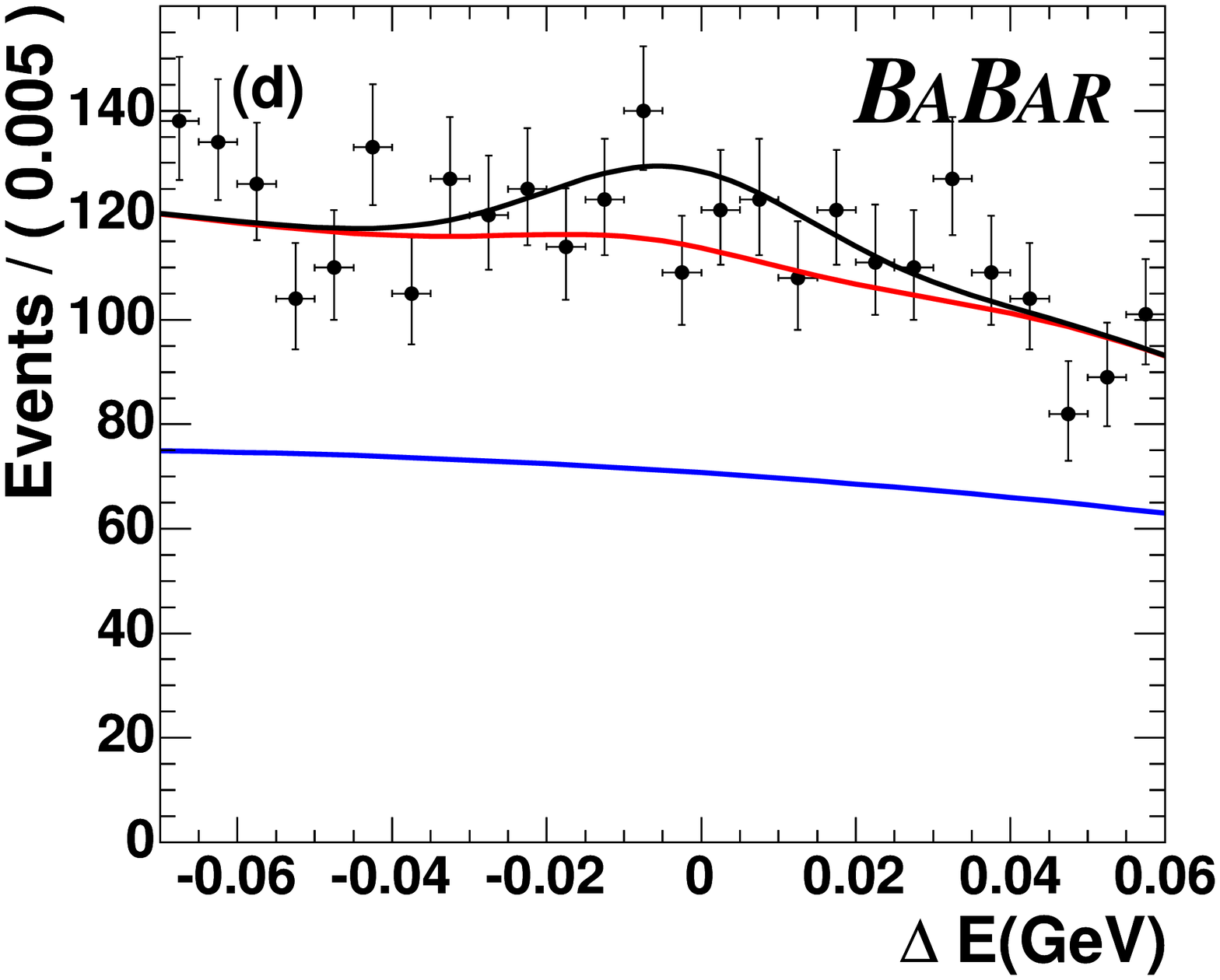} \\
\includegraphics[height=3.5cm]{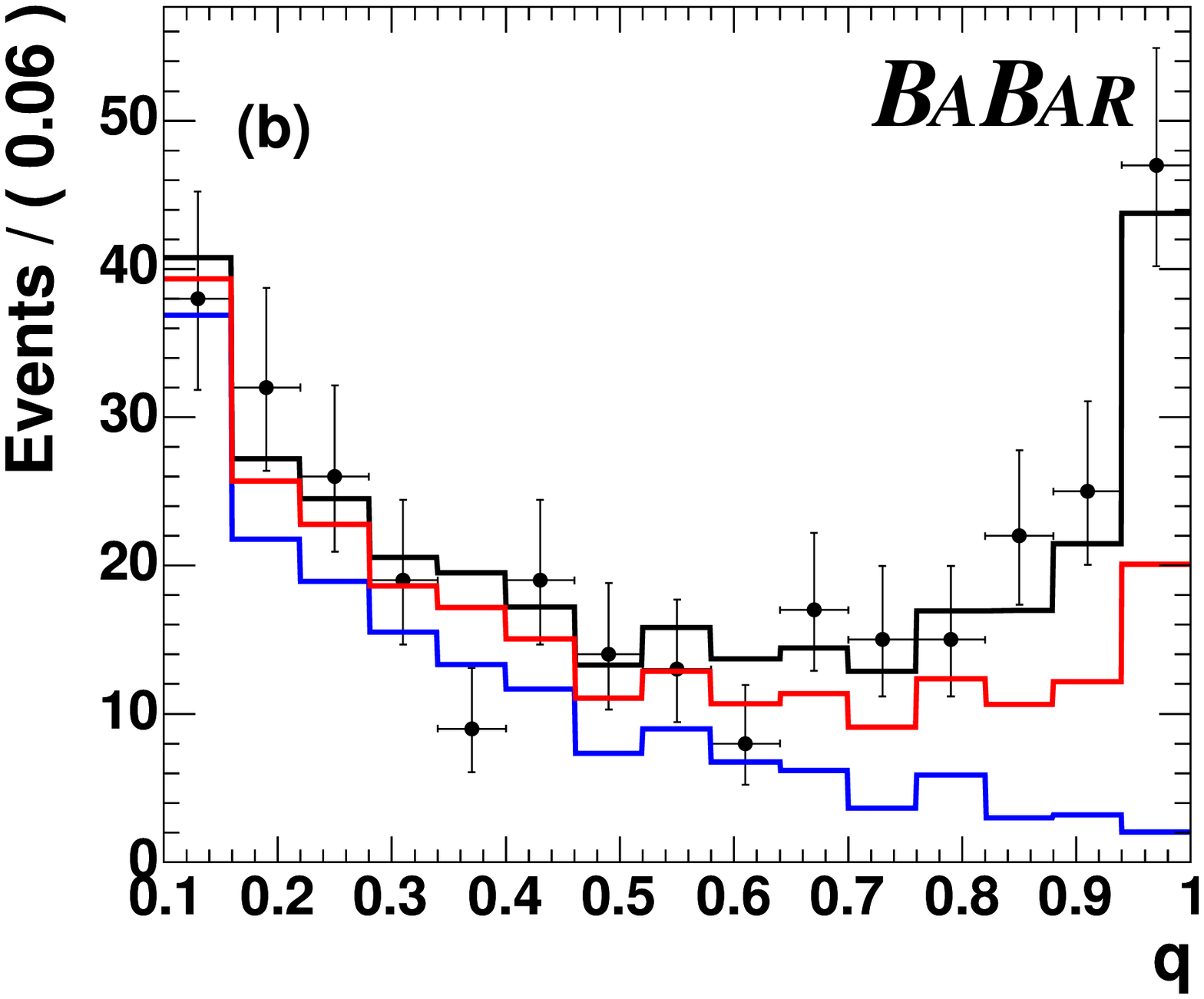} &
\includegraphics[height=3.5cm]{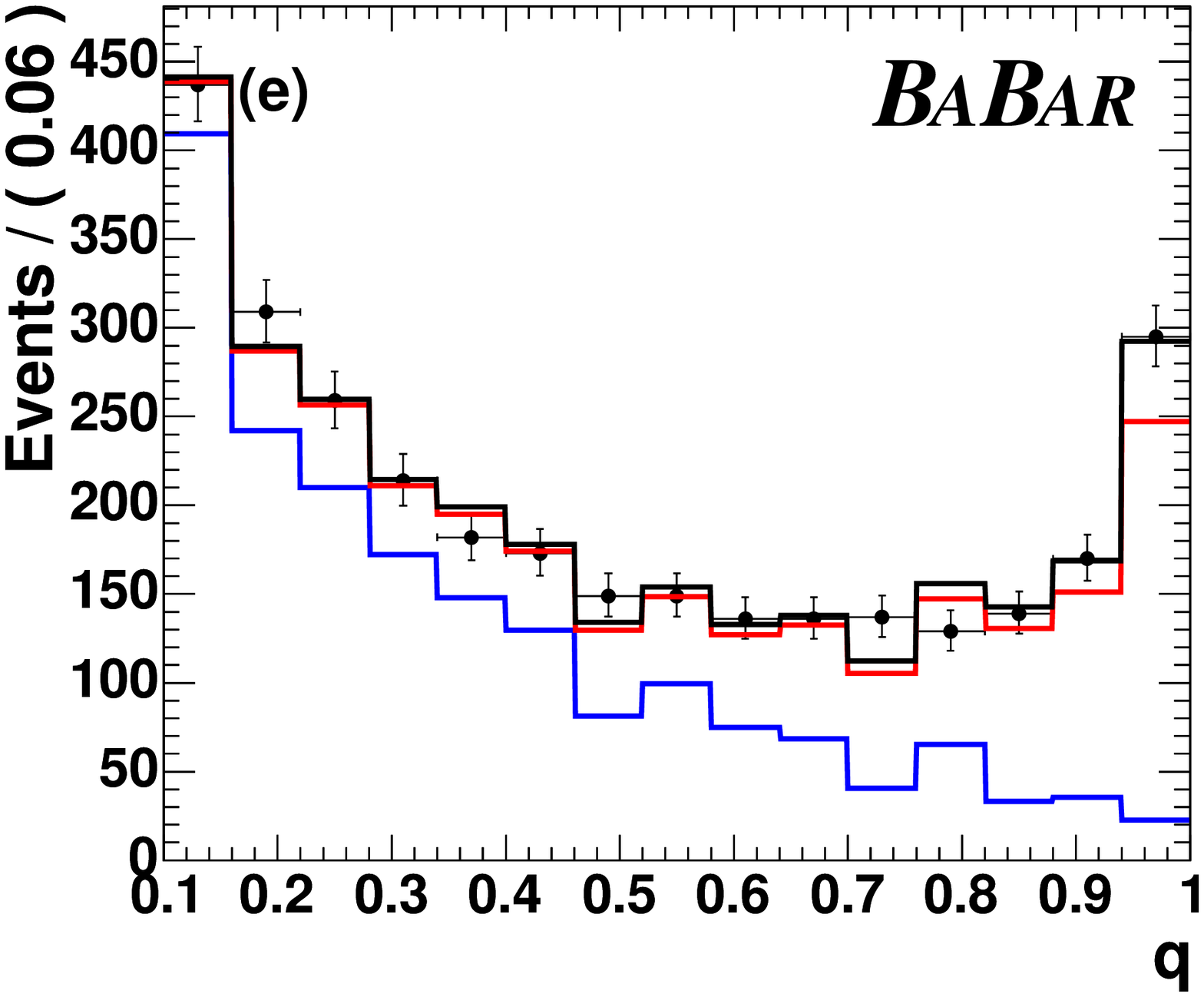} \\
\includegraphics[height=3.5cm]{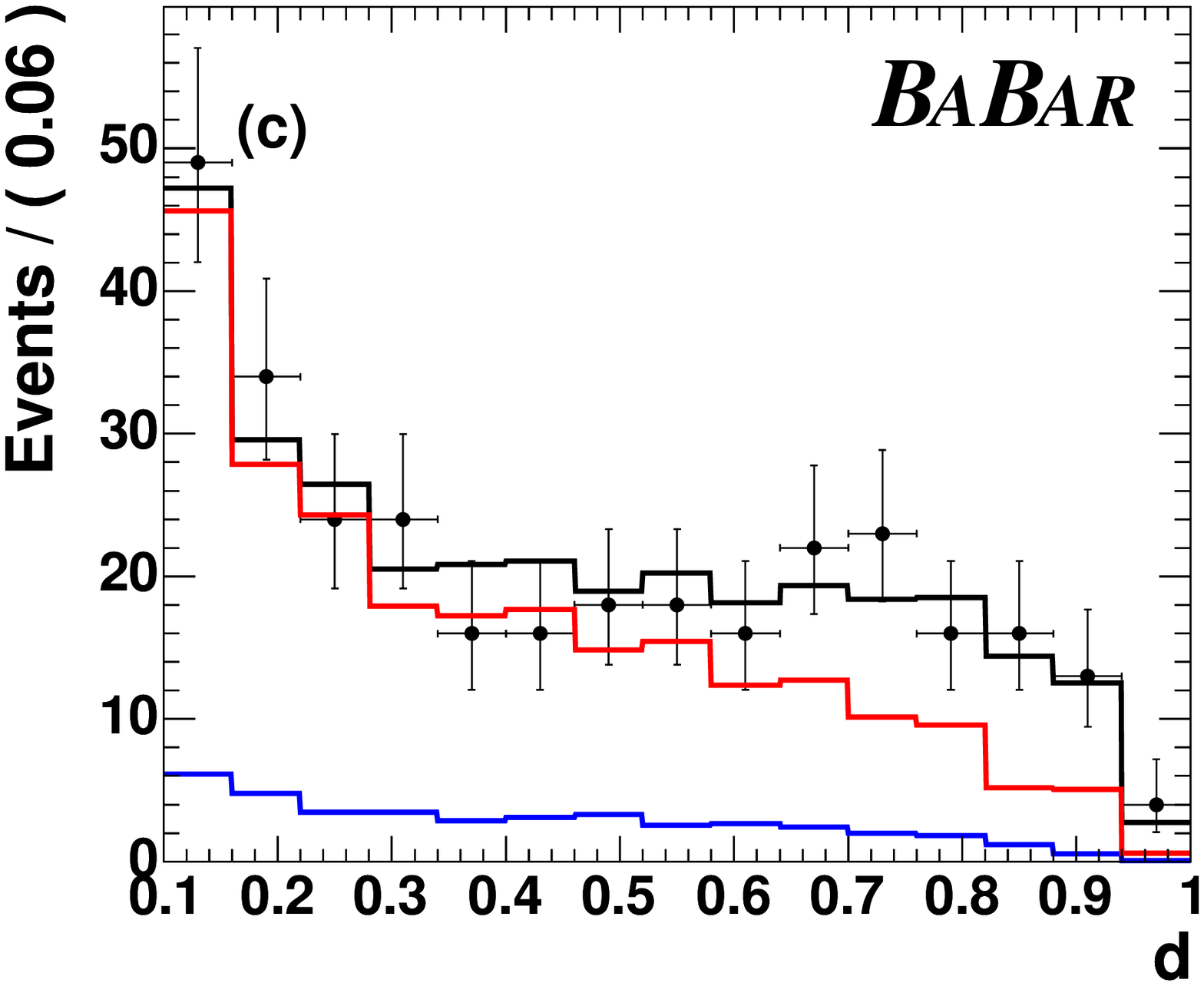} &
\includegraphics[height=3.5cm]{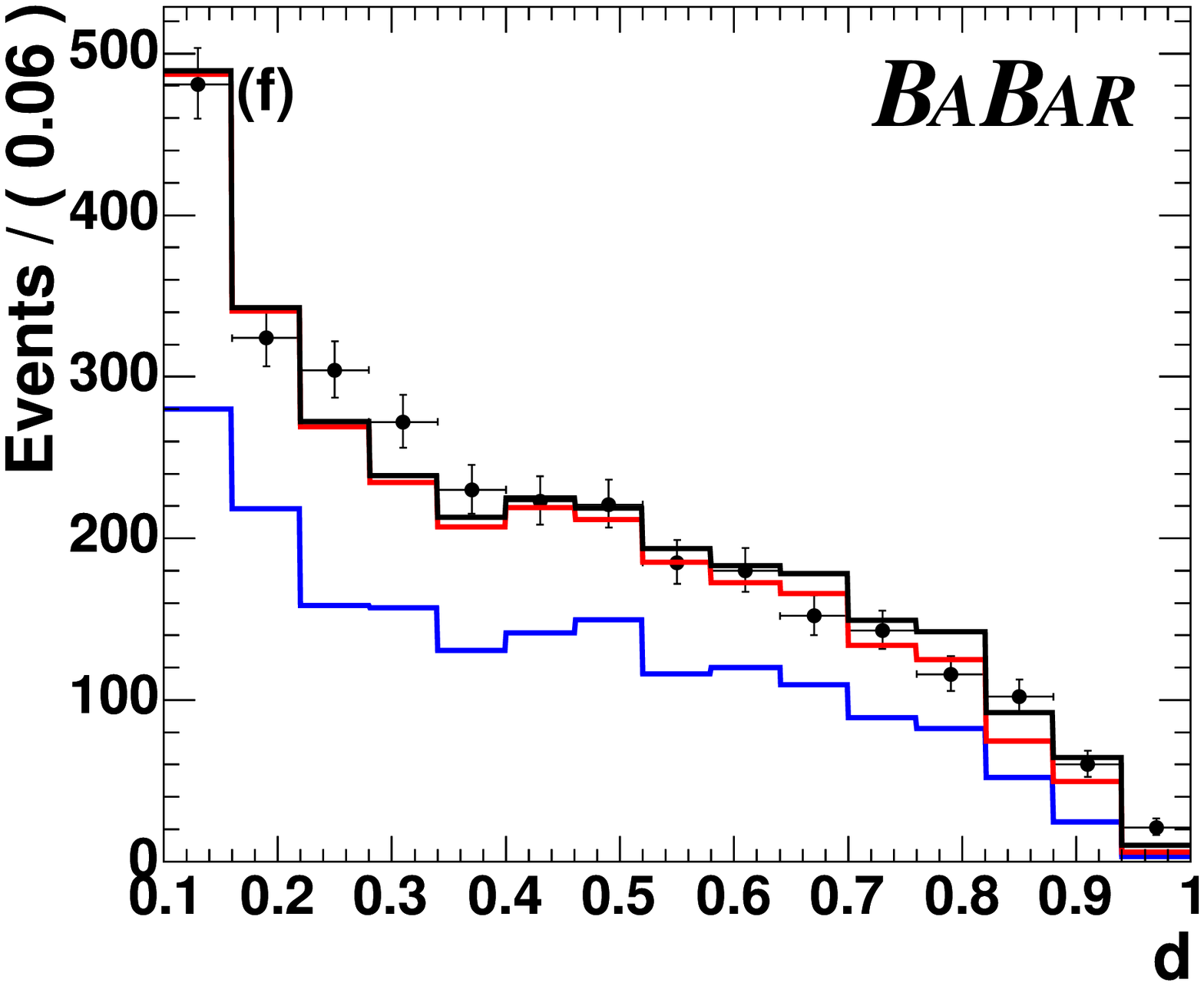} \\
\end{tabular}
\end{center}
\caption{Projections of the data (data points) and fit function 
	onto the (a) \DE, (b) \NNqq,
	and (c) \NNcomb\ axes. 	
	The curves or histograms in each plot show, from bottom
	to top, the cumulative contributions of the continuum, 
	\BB\ background, and signal components of the fit function.
	For each of the variables plotted, a 
	tight event selection is
	applied on the other two variables with a signal efficiency of
	about 50\%, in order to increase the signal-to-background ratio
	in these plots.
	Figures (d), (e), and (f) show the same projections  
	for the entire data sample.}
\label{fig:data-fit-like}
\end{figure}

The fit parameter most correlated with the signal yield is $\NBB$,
with correlation matrix element $\rho(N_\DKgood, \NBB) = -0.33$.
The largest correlation matrix element for the asymmetry is $\rho(\A,
N_\DKgood) = -0.036$.
Projections of the data and the fit function onto the fit variables
are shown in Fig.~\ref{fig:data-fit-like} for events with a high likelihood
of being signal and for the entire data sample.

\begin{table}[!htbp]
\caption{Results of the data fit. Errors are statistical only.}
\begin{center}
\begin{tabular}{lrcl}\hline\hline
Parameter          & \multicolumn{3}{c}{Value}  \\ 
\hline
$N_\DKgood$   &  133    &$\pm$&  23  \\ 
$\A$          &  0.02   &$\pm$&  0.16 \\ 
\hline
$N_\Dpigood$  &  43     &$\pm$&  16  \\
$N_\qqbad$    &  1454   &$\pm$&  53 \\
$\NBB$        &  806    &$\pm$&  54 \\
$R_\Dpix$     &  0.82   &$\pm$&  0.11 \\
\hline\hline
\end{tabular}
\end{center}
\label{tab:results}
\end{table}

The systematic uncertainties in the signal branching fraction and asymmetry 
measurements are
summarized in Table~\ref{tab:syst}. We describe briefly the
procedures used for their evaluation:
(1) The statistical errors in the simulated samples used to obtain
the shapes of $\E_t(\DE)$, $\Q_t(\NNqq)$, and $\C_t(\NNcomb)$ are
propagated to the final fit.
(2) The value of $N_\DKX$ is varied by $\pm 25\%$, 
determined from the uncertainties on the decay modes contributing to 
the \DKX\ background~\cite{ref:pdg}.
The parameters $N_\Dpibad$, $N_\BBgood$, $N_\qqgood$, and $N_\DKbad$
are varied by $\pm 50\%$, which is estimated to be very conservative, given
the level of data-simulation agreement. 
(3) We evaluate the effect of possible differences between the
event distributions in the data and the simulation 
by studying events in the \mes\ sideband $5.23 < \mes < 5.26$~\gevcc,
as well as events produced in the copious decay mode \bMtodpiKpp. 
%
(4) The \DE\ and \NNcomb\ distributions of \DKgood\ events are
slightly correlated, and this correlation is ignored in the PDF.  To
evaluate the uncertainty due to this, we repeat the fit with 
$\C_\DKgood(\NNcomb)$ taken from simulated \DKgood\
events in different $\DE$ bins.
(5) We consider the effect of a possible contribution of charmless 
$B^-\to K^-\pi^+\pi^-\piz$ events, assuming 
a branching fraction of $6\times 10^{-5}$.
(6) The uncertainty in the contribution of a non-resonant component
to the $\dtoppp$ decay~\cite{Frolov:2003jf} is propagated to the
signal efficiency.
(7) We account for the possibility of charge-dependence in the track
reconstruction efficiency.
(8-9) We assign a reconstruction efficiency uncertainty of 1.4\% per charged 
track and 3.5\% for the $\piz$. 
(10) We account for the uncertainty in the number of \BB\ events 
produced by \pep2\
and (11) the uncertainty in the efficiency of the particle
identification requirements applied to the data sample.

\begin{table}[!htbp]
\caption{ Fractional systematic error in the signal branching fraction 
$\B $ and absolute error in the asymmetry $\A$. }
\begin{center}
\begin{tabular}{clcc}\hline\hline
No. &Source             & \multicolumn{2}{c}{Error (\%)}   \\ \cline{3-4}
    &			& $\sigma_\B/\B $ & $\sigma_\A$ \\
\hline
(1)& Simulated sample statistics		&   7.9   & 1.8   \\
(2)& Variation of fixed yields  		&   6.2   & 0.25  \\ 
(3)& Data-simulation shape comparison		&   5.8   & 1.6   \\
(4)& $\DE - \NNcomb$ correlations in $\P_{\DKgood}$&  1.9 & 0.39  \\    
(5)& Charmless branching fraction 		&   0.85  & 0.093 \\
(6)& Dalitz plot distribution  			&   0.33  & \--    \\
(7)& Detector asymmetry  			&   \--    & 0.90   \\
(8)& Track reconstruction efficiency		&   4.2   & \--    \\
(9)& $\piz$ efficiency				&   3.5   & \--    \\
(10)& Number of $\BB$ events produced		&   1.1   & \--    \\
(11)& Particle ID efficiency			&   1.0   & \--    \\
\hline		
  & Total     					& 13	  & 2.6 \\
\hline\hline
\end{tabular}
\end{center}
\label{tab:syst}
\end{table}

Additional cross-checks are performed to verify the validity
of our results.
We compare the fit variable distributions of the data with those of
simulated events in the \DE\ sideband $90 < \DE < 140$~\mev.
The fit variable distributions of simulated continuum events are
validated against the off-resonance data. 
The simulated distributions of \bMtodpippp\ events are
compared with their distributions in the data.
We verify the signal efficiency by measuring the branching fraction
$\B(\bMtodpi)$ using $\Dz$ decays to \dztokpp\ and \dztoppp.
The simulated distributions of the $\NNqq$ and $\NNcomb$ input
variables are compared with the distributions in the data.
In all cases, good agreement between simulation and data is observed.
No significant excess of signal events is found in 
a fit to data events in the \mD\ sidebands $1.775 < \mD < 1.800$~\gevcc
and $1.920 < \mD < 1.955$~\gevcc.
We conduct fits to event samples containing simulated signal and
background events and find no significant biases in all the fit variables.
Fits to parameterized experiments generated with the 
parameter values obtained in the data fit are unbiased, and
their distributions of fit parameter errors and 
maximum likelihood are consistent with those of the data fit.

In summary, using a sample of $229 \pm 2.5$ million $\epem\to\BB$ events 
we observe $133 \pm 23$ 
events in the decay chain \decaychain, where the $\ppp$ final state
excludes the CP-eigenstate $\KS\piz$.
We extract the branching fraction and decay rate asymmetry
\beqa
\B(\decaychain) &=& (5.5 \pm 1.0 \pm 0.7)\times 10^{-6}, \nonumber\\
\A &=& 0.02 \pm 0.16 \pm 0.03,
\eeqa
where the first errors are statistical and the second are systematic.
The level of background suppression we achieve is critical for
using this mode to measure $\gamma$. The
remaining background doubles the statistical error on $\gamma$ with respect to the
no-background case.


\input acknowledgements

\end{document}

%% file: authors_apr2005.tex
%
\author{B.~Aubert}
\author{R.~Barate}
\author{D.~Boutigny}
\author{F.~Couderc}
\author{Y.~Karyotakis}
\author{J.~P.~Lees}
\author{V.~Poireau}
\author{V.~Tisserand}
\author{A.~Zghiche}
\affiliation{Laboratoire de Physique des Particules, F-74941 Annecy-le-Vieux, France }
\author{E.~Grauges}
\affiliation{IFAE, Universitat Autonoma de Barcelona, E-08193 Bellaterra, Barcelona, Spain }
% ...
\collaboration{The \babar\ Collaboration}
\noaffiliation

%% file: acknowledgements.tex
We are grateful for the 
extraordinary contributions of our \pep2\ colleagues in
achieving the excellent luminosity and machine conditions
that have made this work possible.
The success of this project also relies critically on the 
expertise and dedication of the computing organizations that 
support \babar.
The collaborating institutions wish to thank 
SLAC for its support and the kind hospitality extended to them. 
This work is supported by the
US Department of Energy
and National Science Foundation, the
Natural Sciences and Engineering Research Council (Canada),
Institute of High Energy Physics (China), the
Commissariat \`a l'Energie Atomique and
Institut National de Physique Nucl\'eaire et de Physique des Particules
(France), the
Bundesministerium f\"ur Bildung und Forschung and
Deutsche Forschungsgemeinschaft
(Germany), the
Istituto Nazionale di Fisica Nucleare (Italy),
the Foundation for Fundamental Research on Matter (The Netherlands),
the Research Council of Norway, the
Ministry of Science and Technology of the Russian Federation, and the
Particle Physics and Astronomy Research Council (United Kingdom). 
Individuals have received support from 
CONACyT (Mexico),
the A. P. Sloan Foundation, 
the Research Corporation,
and the Alexander von Humboldt Foundation.